%% file: paper_main.tex
\documentclass[conference]{IEEEtran}
\IEEEoverridecommandlockouts
\usepackage{cite}
\usepackage{amsmath,amssymb,amsfonts}
\newtheorem{theorem}{Theorem}

\newtheorem{corollary}{Corollary}
\usepackage{algorithm} \usepackage{algpseudocode}

\usepackage{graphicx}
\graphicspath{ {./figures/} }
\usepackage{textcomp}
\usepackage{xcolor}
\def\BibTeX{{\rm B\kern-.05em{\sc i\kern-.025em b}\kern-.08em
    T\kern-.1667em\lower.7ex\hbox{E}\kern-.125emX}}

\algnewcommand\INPUT{\item[\textbf{Input:}]}%
\algnewcommand\OUTPUT{\item[\textbf{Output:}]}%
\newcommand{\pluseq}{\mathrel{+}=}

\newcommand{\upperRomannumeral}[1]{\uppercase\expandafter{\romannumeral#1}}

\usepackage{lettrine}
\newcommand{\RNum}[1]{\uppercase\expandafter{\romannumeral #1\relax}}

\begin{document}

\title{{\huge On the Convergence Rates of Learning-based Signature\\ Generation Schemes to Contain Self-propagating Malware}\\
\thanks{This work was funded by NSF grant CNS-1413996 ``MACS: A Modular Approach to Cloud
	Security.''}
}
\author{\IEEEauthorblockN{Saeed Valizadeh}
\IEEEauthorblockA{\textit{Dept. Computer Science and Engineering} \\
\textit{University of Connecticut}\\
Storrs, CT, USA \\
saeed.val@uconn.edu}
\and
\IEEEauthorblockN{Marten van Dijk}
\IEEEauthorblockA{\textit{Dept. Electrical and Computer Engineering} \\
\textit{University of Connecticut}\\
Storrs, CT, USA \\
marten.van\_dijk@uconn.edu }
}

\maketitle

\begin{abstract}
In this paper, we investigate the importance of a defense system's learning rates to fight against the self-propagating class of malware such as worms and bots. To this end, we introduce a new propagation model based on the interactions between an adversary (and its agents) who wishes to construct a zombie army of a specific size, and a defender taking advantage of standard security tools and technologies such as honeypots (HPs) and intrusion detection and prevention systems (IDPSes) in the network environment.  As time goes on, the defender can incrementally learn from the collected/observed attack samples (e.g., malware payloads), and therefore being able to generate attack signatures. The generated signatures then are used for filtering next attack traffic and thus containing the attacker's progress in its malware propagation mission. Using simulation and numerical analysis, we evaluate the efficacy of signature generation algorithms and in general any learning-based scheme in bringing an adversary's maneuvering in the environment to a halt as an adversarial containment strategy.
\end{abstract}

\begin{IEEEkeywords}
Botnet, Malware propagation modeling, Self-replicating code,  Worms, Intrusion detection and prevention system, Honeypots, Security games
\end{IEEEkeywords}

\section{Introduction} \label{sec:intro} 
\input{sections/intro}

\section{Background and Related Work}\label{sec:backg} 
\input{sections/bg_new}

\section{A Learning-based Propagation Model}\label{sec:model}
\input{sections/DE}

\section{Numerical Analysis of the Learning-based Model}
\input{sections/num_analysis}\label{sec:numAna}

\section{Simulation Results}\label{sec:sim}
\input{sections/sim}

\section{Concluding remarks and future directions} \label{sec:con} 

\input{sections/conclusion}

\bibliographystyle{IEEEtran}
\bibliography{sections/references}

\appendix
\input{sections/appendix}\label{appendix}

\end{document}

%% file: sections/intro.tex
With the rise of advanced and novel malware (e.g., Stuxnet worm \cite{jin2018snapshotter}, and WannaCry ransomware cryptoworm \cite{cisco:Misc}) and the expanding population of Internet residents, especially the explosion of connected Internet of Things (IoT) devices, new methods, and research for understanding malware behaviors and jousting with sophisticated botnets seem to be crucial.
Of specific interest to us are two types of malware, worms \cite{weaver2003taxonomy}, and autonomous bots \cite{complexbot:Misc}, mainly  due to their wide-ranging audience, and self-propagating characteristics. These types of malware execute a ``seek-and-infect'' mission strategy. For instance, a scanning worm conventionally infects a target by utilizing vulnerability exploitation, and it automatically probes the environment to \emph{independently} propagate itself (or modified copies of itself) from the infected machine to other vulnerable hosts in a network. Consequently, a straightforward way of assembling an army of infected hosts is spreading the malicious code in the form of a scanning worm. This zombie army which is also known as a botnet is indeed a collection of compromised machines distributed over a network (usually the Internet) and controlled by its originator (known as a botmaster) through a command and control (C\&C) structure \cite{feily2009survey}. 
Afterward, these infiltrated workstations (bots) can be managed for different malicious purposes including but not limited to ransom campaigns, massive spam-marketing, large-scale information harvesting and processing, click fraud, pay-per installation, and, distributed denial-of-service (DDoS) attacks.

The never-ending battle between adversaries and system defenders drives both parties to improve their methods and technologies to enhance their chances of prosperity. On the one hand, attackers and network penetrators come up with new attack methods and evasion techniques to bypass (or even deny) defensive systems in any possible manner. Code obfuscation, session splicing, fragmentation attacks, string matching and DDoS attacks \cite{marpaung2012survey} are among the most common techniques used by the adversaries to decrease the attack detection probability or to increase the chances of attack prosperity.
An armored malware (e.g., a metamorphic or polymorphic worm) buys itself more propagation time by being hard to identify and disassemble. 
On the other hand,  a large number of studies on malware detection/prevention techniques, IDPSes, and HP technologies have been carried out to present more effective methods and designs to reach better accuracy and performance \cite{idika2007survey,bhuyan2014network,chen2006survey}.
For instance, fast and automatic signature generation schemes are introduced to fight against even the mightiest type of malware \cite{kim2004autograph,newsome2005polygraph,kreibich2004honeycomb,singh2004automated,li2006hamsa,tang2011signature,wang2010thwarting}. Some claim that they can generate high-quality signatures (with negligible false alarm rates) based on only a few malware samples\footnote{E.g., see Hamsa \cite{li2006hamsa} which only requires 100-500 malware samples in the suspicious flow pool to reach a \%5 false negative rate, even for a previously unknown worm} usually captured via honeypots or a heuristic flow classifier.

Although there exist various number of research for worm-like malware propagation modeling, in addition to proposals for containment strategies such as content-based filtering based on automatic and distributed signature generation of the malware, the question unanswered here is \emph{how and why does the malware spread, especially the kinds targeting a broad audience, if it is the case that we can generate high quality and reliable signatures, even for an armored malware?}

\noindent
{\bf Our Approach.} 
To answer such questions, the effectiveness of content-based filtering strategies\footnote{As one of the most effective and feasible solutions to tackle the spreading of worm-like malware} based on automatic signature generation schemes must be studied. Hence, in this paper, instead of presenting yet another malware detection/prevention methodology, we revisit the spread of malware phenomenon, especially the self-propagating ones such as worms and bots, from a new perspective. We introduce a new model for capturing the interactions of an adversary and its agents with the defensive system during the early phases of a zombie army construction. 
Unlike a malware propagator who usually follows a passive attack strategy for malware distribution to a broad audience, meaning that the attack technology usually remains the same after unleashing the first few samples of the malware, the defender can incrementally learn regarding the attack technology (specifically, from observed malware payloads). As time elapses, and the defender captures more attack data, it can reach better detection rates and accuracy. This learning rate indeed reflects into higher quality malware signatures which can be used for online malicious traffic filtering and hence malware containment purposes.

\noindent
{\bf Contributions \& Results.} The main contributions of our work can be summarized as follows:

\begin{itemize}
\item Motivated by the advances in networking infrastructure and technologies such as the introduction of software-defined networking (SDN), and also the widespread embrace of automated machine learning (AutoML) techniques in different realms of networking and security fields, we introduce a novel malware propagation model called ``learning-based model'' suitable for today's technologies and more advanced malware. 


\item The previous outdated models are mainly focused on the attacker's strategies (specifically target discovery and scanning rates \cite{chen2009information,sellke2008modeling,rohloff2005stochastic}), or in best cases, an ill-defined cleaning of the infection or patching processes \cite{zou2002code,chen2003modeling,dagon2006modeling}. However, we show how the increased knowledge of the defender can be taken into consideration by modeling the learning rate of the defense system as a function $f(l)$ (which is the probability of detecting and filtering a next malicious payload) where $l$ represents the number of malicious payloads for a specific exploit so far collected.

\item The model leads to the development of an automatic self-replicating malware containment strategy that prevents the spread of the malware beyond its early stages of propagation.
\item We provide precise conditions on the convergence rates of a learning-based signature generation algorithm that determines whether the malware spread will ultimately settle or not.
\item We derive tight upper and lower bounds on the total number of hosts the malware propagator can infect.
\item The model is general enough and enables us to evaluate the effectiveness of learning-based signature generation schemes and containment strategies.
\item We study learning functions of the form $f(l)=(\frac{A}{A+l})^{\alpha}$ in which $A$ is considered to slow down the learning process, and $\alpha$ is the amplification factor in the learning. This enables us to capture characteristics of both yesteryears malware (e.g., a monomorphic worm) and also today's more advanced ones such as polymorphic malware. 
\item Numerical analysis and simulation results show that regardless of attacker's scanning rates, with a proper learning function that converges fast enough to $1$, the propagation will be contained and only a negligible number of susceptible population will be infected.
\end{itemize}

\noindent
{\bf Paper Organization.}
The rest of this paper is organized as follows.
Section \ref{sec:backg} briefly reviews existing worm-like malware's propagation models in addition to automatic signature generation schemes and content-based filtering of such malware. 
We present a new learning-based propagation model in section \ref{sec:model} followed by the corresponding numerical analysis and simulation results of this model for various learning functions in section \ref{sec:numAna}, and \ref{sec:sim}. We finally conclude the paper and discuss future works in section \ref{sec:con}. 

%% file: sections/bg_new.tex
The three potential solutions to mitigate a self-replicating code's rampancy are prevention, treatment, and containment strategies. The prevention and treatment schemes can be summarized as following secure software/application development procedures in hopes of having vulnerability-free software, or patching/updating vulnerable systems as soon as vulnerabilities are discovered. Unlike the containment techniques, such conducts are usually vital for pre/post incident time and not during the period of an incident. In this section, we first review content-based filtering and automatic signature generation schemes as the most effective containment strategy \cite{moore2003internet} to hinder the spread of a malware. Moreover, we study some of the most relevant propagation models existing in the literature before presenting our model.

\subsection{Content-based filtering and Automatic Signature Generation Schemes}

Human interventions and attack response time must be minimized in order to be able to prevent widespread infections. To this end, automatic, distributed and real-time detection and containment strategies, even for the mightiest type of worm, i.e., a polymorphic one, are introduced in the literature. Such signature-based detection and prevention techniques are of particular interest of this work as ``an ounce of prevention is worth a pound of cure'' meaning that such containment strategies can be very efficient for thwarting the spread of malware, especially during the early phases of a botnet construction and worm propagation, only if a high-quality signature can be generated (automatically).  Such signatures can be created based on various attributes (e.g., length of the fields or invariant substrings of the byte sequences) of a class of malware, and later these signatures will be used for content-based malicious traffic filtering (even for extreme cases such as previously unknown malware or an armored one, e.g., a polymorphic worm). While designing a new automatic signature generation scheme is not the objective of this manuscript, here we review the most notable works in this area in order to better understand such schemes especially their architecture so that we can build a realistic propagation model based on a general learning algorithm (on which the signature generator is built upon) which we explain in later sections.

Autograph \cite{kim2004autograph} is one of the first proposals for automatic (and optionally distributed) signature generation for a polymorphic worm, utilizing a naive portscan-based flow classifier (for TCP worms) to lessen the volume of traffic on which it performs a content-prevalence analysis. The flow classification enables Autograph to construct a suspicious flow pool on which it executes TCP flow reassembly for the payloads and outputs the most frequently occurring byte sequences across the flows as signatures. EarlyBird \cite{singh2004automated} uses a similar approach as Autograph in generating signatures by taking advantage of Rabin fingerprints.

On the other hand, Polygraph \cite{newsome2005polygraph} forms signatures that consist of multiple disjoint content substrings to address the inefficiency of single, contiguous string-based signature generation techniques (such as Honeycomb \cite{kreibich2004honeycomb}, EarlyBird, and Autograph). 
Its underlying assumption for signature generation is that for a real-world exploit to function correctly, various invariant substrings must often be present in all alternatives of a malware payload; and these substrings typically correspond to return addresses, protocol framing, and in some cases, defectively obfuscated code. Hamsa \cite{li2006hamsa} is another fast content-based signature generation method which generates multiset tokens as signatures. In comparison with Polygraph, it can provide better attack resilience and noise tolerance (accuracy). PolyTree \cite{tang2011signature} can show how the worm variants evolve and make the signature refinement task upon a new worm sample arrival quick using an incremental signature tree construction. This is based on the observation that worm signatures are related and a tree structure can properly reflect their familial resemblance which enables organizing the extracted signatures from worm samples into a tree structure. Unlike the aforementioned proposals which generate \emph{exploit-specific} signatures, the paper in \cite{wang2010thwarting} presented an automatic \emph{vulnerability-driven} network-based length-based signature generator called LESG for zero-day polymorphic worms exploiting buffer overflow vulnerabilities based on the fact that specific protocol fields in such attacks are usually longer than those in a conventional protocol usage.


\subsection{ Propagation Modeling and Estimating a Botnet Size}\label{bg.malProModels}

In malware propagation models, possible states concerning vulnerable population during a malware attack are susceptible (S), infectious (I), and recovered (R). According to a host's state at different times, and based on the transition between such states, malware propagation models can be categorized into three classes: if a host can only have one of the susceptible or infectious states and not being able to be recovered after an infection, the model is called susceptible-infectious (SI). If the model considers a permanent revival for an infected machine, meaning that the host remains in an immune state after the recovery process, the model is called susceptible-infectious-recovered (SIR) and finally if there is the chance of being infected again after a recovery the model is called susceptible-infectious-susceptible (SIS). 

The classical simple epidemic model which is an SI modeling is the most commonly used model in the literature and can be simply described by the following differential equation in which $k$ denotes the total number of vulnerable nodes to the epidemic (i.e., the susceptible population), and at each time step, the so far infectious hosts $i(t)$ propagate the epidemic with a constant rate $\beta$.  

\begin{equation}\label{eq.SEM}
\frac{d i(t)}{dt} = \beta i(t)[ k - i(t)] 
\end{equation}

An extension to the classical simple epidemic model is the Kermack-McKendrick model (also known as the classical general epidemic model \cite{roberts2004mathematical}) which is one of the mostly used SIR models in the literature. In this modeling, the effects of a removal process in the infectious hosts population is taken into consideration. Consider $r(t)$ as the number of removed hosts from previously infected hosts, and $\zeta$ as the constant rate of removal, hence, the Kermack-McKendrick model can be expressed by:
\begin{equation}\label{eq.KerMc}
\frac{d i(t)}{dt} = \beta i(t) [k - i(t) - r(t)]  - \frac{d r(t)}{dt}
\end{equation}
where $\frac{d r(t)}{dt} = \zeta i(t)$.

Another proposal based on Kermack-McKendrick model is \cite{dagon2006modeling} in which it provides a diurnal model for different time zones for botnet propagation modeling by introducing a correction factor $\alpha(t)$ (``the diurnal shaping function'') in the total number of online (available) hosts based on the time region. Therefore the diurnal worm propagation model can be represented by:
\begin{equation}
\frac{d i(t)}{dt} = \beta \alpha^2(t) i(t)[ k(t) - i(t) - r(t)] - \zeta \alpha(t) i(t)
\end{equation}
The two-factor model \cite{zou2002code} is as an extension to Kermack-McKendrick's model which enhances it in two manners. First, it separates the removal process into two parts, one for the elimination of infectious hosts (due to cleaning) and the other for purging the susceptible population (due to patching and system updates). Second, it considers a time-varying infection rate to take worm traffic's impact on the network infrastructure into consideration (e.g., congestion in the network). 

\begin{equation}\label{eq.2fact}
\frac{d i(t)}{dt} = \beta(t) i(t) [ k(t) - i(t) - (r(t)+q(t))] - \frac{d r(t)}{dt}
\end{equation}
where $\frac{d r(t)}{dt} = \zeta i(t)$, and $q(t)$ denotes the number of removed hosts from the susceptible population.


Analytical active worm propagation model (AAWP) which is a similar approach in the discrete time setting is presented in \cite{chen2003modeling}.

Stochastic modeling of active worms is another line of research in this area. Sellke \textit{et al.} \cite{sellke2008modeling} model the propagation of the malware through a branching process, i.e., each infected machine in one generation will independently produce some random number of infected machines in next-generation according to a fixed probability distribution. The model can determine the extinction condition of the malware and provide an upper bound on the total number of infected hosts. Rohloff \textit{et al.} \cite{rohloff2005stochastic} used a simple stochastic density-dependent Markov jump process to model worm propagation. Each state in the chain represents the number of so far infected machines $i(t)$, in addition to the total number of remaining susceptible population $s(t)$. The model does not consider any removal/cleaning actions meaning that the total number of vulnerable hosts $k = i(t) + s(t)$ is always constant at each transition of the Markov chain. The time it takes for this chain to reach its absorbing state is calculated. For more information regarding malware propagation models we refer to the survey presented in \cite{wang2014modeling}.


The models mentioned above mainly focus on the treatment processes --such as patching a susceptible node or cleaning an infected machine-- for vulnerable population and infection rate reduction purposes. 
In practice, these treatment strategies, however, are not suitable for the expeditious spread of the malware and therefore short-term reliefs for an outbreak. Moore \textit{et al.} \cite{moore2003internet} showed that a content-based filtering strategy is the most vital containment strategy to limit the spread of the malware via isolating it from the susceptible population. Therefore, in this paper, we mainly focus on the containment solutions made possible based on a defense system's learning engine which makes the propagation of the epidemic more difficult. Valizadeh and van Dijk \cite{valizadeh2019toward} originated the notion of learning in Markov-based cyber-attack modeling and described a general ``game of consequences'' in which the attacker's chances of making a progressive move in the game depends on its previous actions. This notion of learning has been previously used in learning-based signature generation schemes for the polymorphic worm but never been applied to propagation models. The advances in network technologies (especially SDN) and automated machine learning (AutoML) motivate us to take advantage of such learning mechanisms and introduce a new \emph{learning-based} propagation modeling suitable for today's and future's infrastructure and technology.



Moreover, the most critical shortcoming of the traditional models is that they suffer from not taking \emph{both parties}' capabilities, actions and strategies into consideration for the model development. Therefore, these \emph{static} models are incapable of representing the \emph{dynamic} nature of today's network attack-defense scenarios. Our modeling differs from the previously mentioned models in the sense that as the time elapses, we can take enhanced attack/defense strategies into consideration by recognizing the players' interactions as a learning process, especially from the defense systems' perspective. This presented framework enables us to explicitly study how the learning rates of a defense system can affect future interplays of the players and their probability of success in the malware propagation game. Moreover, instead of considering an ill-defined removal/cleaning rate, we can describe where the containment process could come from by considering an incremental learning mechanism for the defender.

%% file: sections/DE.tex
In this section, we study how a content-based filtering strategy--which utilizes a general learning-based signature generation scheme-- can play a role in containing the propagation of a self-replicating code. More specifically, regardless of the limits on the \emph{accuracy} of any learning-based automatic signature generation algorithm, we study the \emph{efficacy} of such schemes in ceasing the propagation of the malware under the assumption of their constructability.

To understand the effect of a defense system's learning engine on the containment of the malware, we model the defender's learning engine (i.e., a signature generator algorithm) as a function $f$, which takes so far collected/observed suspicious traffic $l$ (i.e., attack payloads) and outputs a signature for that class of malware. This generated signature is then used for online content-based traffic filtering in which each incoming malware traffic can be filtered with probability $f(l)$ (true positive) or with $1-f(l)$ the system fails in detecting a malicious packet (or it falsely labels it as benign, i.e., false negative). Later,  we briefly discuss the occurrence of false positives in the model. Moreover, we are interested in the infection modeling during the Window of Vulnerability (WoV) time as the number of vulnerable systems is not yet shrunken to insignificance and the attacker’s exploit is useful in this period. This means that we do not take patching and cleaning of the susceptible and infected population into consideration.

\subsection{Idealized Deployment, Network and Learning Model}
We consider a logically centralized defense system (e.g., an IDPS) employed in the network, that is the containment system is universally deployed within the address space, and the learning engine works with all the observed/collected information regarding the worm infections (i.e., samples) and then distribute the generated signatures to defense system agents (e.g., edge routers, inline network-based intrusion detection systems (NIDS)). Similar to \cite{venkataraman2008limits}, we also consider that the learning algorithm will update its internal state after observing each new incoming batch of data, and these updates will be made over all of the so far accumulated samples. In this regard, with these updates, the learning engine might be able to find a high-quality signature over a more extended period, while without such updates, no learning would be possible. Moreover, we assume that the signature generation task on the accumulated samples is performed with no delay. Also, the signature distribution to the defense agents will be done immediately.

Automatic signature generation schemes (e.g., see \cite{kim2004autograph,newsome2005polygraph,li2006hamsa,tang2011signature}) commonly take advantage of a heuristic flow classifier for constructing a suspicious flow pool in order to reduce the volume of traffic on which further analysis must be performed. Therefore, we also consider that a copy of traffic is given to a flow classifier for suspicious flow pool construction. For simplicity and generality, however, we model the defense system's classifier and traffic analysis task as a probabilistic sampling process \cite{sperotto2010overview} in which each malicious packet can be sampled with probability $\lambda$. Note that $\lambda$ reflects the classifier's accuracy in which it is the probability that an incoming packet will be marked as suspicious traffic conditioned on the fact that it is indeed malicious. The captured payloads from the classifier are given to the signature generator engine for signature generation. The generated signatures then are used for the inline packet filtering by the defense system. This allows us not to be concerned about the transmission protocol (TCP/UDP), and in general more stealthier propagation schemes (e.g., second channel delivery) which are usually more robust against anomaly-based detection systems, as they will not trigger any events during the propagation\footnote{Note that this is a practical assumption as the most common form of a self-replicating code found in the wild is a \emph{self-carried} (malware payload is transferred in a packet by itself) malware which, regardless of the transmission scheme (TCP or UDP) utilizes a blind scan strategy as its target discovery (see Table 1 in \cite{li2008survey}).}. 

One very common problem among any IDPS is the inability to provide absolutely complete and accurate detection rates. This incompleteness and inaccuracy usually lead to the occurrence of a false detection of a malicious activity as benign or vice versa. 
Due to the existence of false alarms (especially false positives), it is a common practice not to utilize black/whitelisting prevention policies for the IDPS. This means that we do not care about \emph{stealthy} attacks in which the attacker can conceal its real identity and disguise the source of attack traffic to decrease the chance of being located through common known practices such as IP address spoofing, use of proxies, etc. 
However, false positives (labeling a benign packet/activity as suspicious on the classifier level or malicious for inline filtering) play an essential role in the accuracy of the generated signatures and therefore the normal operations of the network in which the defense system is implemented. Note that in our model, false positives can occur on two levels. First, the classifier may mark a benign packet as suspicious which eventually could lead to inaccurate signatures or taking a long time to generate true signatures. To address this problem, we consider a deceleration factor in our modeling to take the impact of such false positives in delaying the signature generation into consideration. Second, since the generated signatures are used for online filtering of incoming packets, a false positive at this level may drop a benign packet at the network level. However, notice that such incidents do not impact the accuracy of our model, i.e., the total number of infected nodes at any instance of time will not be affected.

\subsection{A Learning-based Malware Propagation and Containment Model}

\begin{table*}[!t]
	\renewcommand{\arraystretch}{1.3}
	\caption{Notations in this paper}
	\label{table_notations}
	\centering
	\begin{tabular}{ |p{2cm}|p{10cm}|  }
		\hline
		Notation & Explanation\\
		\hline
		$n$    & Total number of nodes in the address space\\
		$k$  & Total number of vulnerable nodes (i.e., susceptible population)\\
		$\eta$ & Average scanning rate of an infected machine\\
		$i(t)$ & Total number of infected nodes at time $t$\\
		$f(l)$  & Defender's filtering probability based on its current knowledge of malware\\
		$\lambda$  & Probability of marking an incoming malicious traffic as suspicious on classifier level\\
		$j(t)$  & Total number of attack samples collected at time $t$\\
		$\gamma$  & Average rate of attack sample collection by the defense system\\
		$\alpha$  & Amplification factor in learning\\
		$A$  & Deceleration factor in learning \\
		\hline
	\end{tabular}
\end{table*}

Our goal is to show that a sufficiently increasing learning rate will stop malware from propagating and only a few number of susceptible nodes $k$ will be infected. Consider the SI modeling represented in (\ref{eq.SEM}), as follows 
$$ \frac{d i(t)}{dt} = \eta \frac{k-i(t)}{n} i(t),$$
where $i(t)$ represents the number of infected nodes at time $t$, $n$ is the size of address space, and $\eta$ is equal to the average scanning rate of an infected machine. Rather than modeling the above equation, let us assume that we already will be able to bound $i(t)$ to a number $\ll k$ so that
\begin{equation}
 \frac{d i(t)}{dt} = p\cdot  i(t) \label{eqwc}
 \end{equation}
would be a good approximation for some constant 
$$ p = \eta k / n.$$
In fact this approximation gives the adversary an advantage since $\eta \frac{k-i(t)}{n} i(t)\leq p i(t)$ which makes the new differential equation (\ref{eqwc}) a best case scenario for the adversary.

We adapt (\ref{eqwc}) to include prevention of infection: If the defender gathers samples of malicious payloads at a certain rate $\lambda$, then the defender collects $\lambda \cdot \eta \cdot i(t)$ samples at time $t$. Notice that unlike the previous models in which the higher the scanning rate of the malware, the sooner the whole population become infected, in our modeling, however, a high scanning rate will also potentially provide more samples to the defense system which can lead to constructing a valid signature sooner and therefore holding the attack's progress. If we denote the number of samples collected up to time $t$ by $j(t)$, and defining $\gamma = \lambda \cdot \eta$, then $j(t)$ satisfies
\begin{equation} \frac{d j(t)}{dt} = \gamma \cdot i(t). \label{eqj} \end{equation}
The defender's knowledge of how to recognize and prevent malicious payloads increases as a result of observing and collecting malicious traffic. We assume $f(l)$ captures this learning -- with probability $f(l)$, where $l=j(t)$ is the number of samples collected so far, the adversarial payload is prevented from doing any harm. With probability $1-f(l)=1-f(j(t))$, the payload may proceed and this leads to the following adjustment of (\ref{eqwc}):
\begin{equation} \frac{d i(t)}{dt} = p\cdot  i(t) \cdot (1-f(j(t))). \label{eqi1} \end{equation}
Equation (\ref{eqi1}) is the learning-based model and this is what we study; It shows that in addition to hitting a vulnerable target, an infected machine's endeavors to infecting a new device by submitting an ominous packet to the target must not get filtered by the defense system.

We will analyze monotonically increasing learning rates of the form
$$ 1- f(l) = \left( \frac{A}{l+A} \right)^\alpha$$
for some exponent $\alpha$ and constant $A$. The exponent $\alpha$ reflects how collected samples amplify the learning rate while constant $A$ is included to slow down the learning process, i.e., delayed learning. 
In this fashion, the learning function can be adjusted to take both yesteryears' worms and today's more advanced malware's characteristics into consideration in the model. For instance, for a naive, monomorphic malware, which the signature generation is more straightforward and can be done based on only a few malware samples, a small slow-down factor $A$ and a large enough amplification factor $\alpha$ can be used in the model to capture the swiftness of signature generation task. On the other hand, for a more potent malware, e.g., polymorphic/metamorphic worms, a large slow-down factor and small amplification factor can depict the hardness of signature generation task when dealing with such malware. 

Substituting $1-f(l)$ into (\ref{eqi1}) yields
\begin{equation} \frac{d i(t)}{dt} = p\cdot  i(t) \cdot \left( \frac{A}{j(t)+A} \right)^\alpha. \label{eqi2} \end{equation}

Now notice that (\ref{eqj}) expresses $i(t)$ as $i(t) = j'(t)/\gamma$ where $j'(t)=\frac{d j(t)}{dt}$ is the first derivative of $j(t)$. By denoting $j''(t)$ the second derivative of $j(t)$ and substituting these  into (\ref{eqi2}) gives the final differential equation which we wish to solve:
\begin{equation} j''(t)/\gamma = p\cdot  j'(t)/\gamma \cdot \left( \frac{A}{j(t)+A} \right)^\alpha. \label{eqi3} \end{equation}
We are interested in bounding $j'(t)/\gamma = i(t)$. Initially,
\begin{eqnarray*}
j(0) &=& 0 \mbox{ and } \\
j'(0)&=&\gamma \cdot i(0) = \gamma,
\end{eqnarray*}
 meaning that there is a single infected node within the network at time zero (patient zero) and the defender's knowledge of the attack is zero at the beginning (zero-day vulnerability/exploit).

We now present the following theorem (for the analysis and the proof see the appendix) which provides lower and upper bounds on the total number of nodes the attacker can infect.
 \begin{theorem}\label{theo}
 For $\alpha<1$, $i(t)= \Omega(t^{1-\alpha})$. For $\alpha=1$, $i(t)=\Omega(\ln t)$. For $\alpha >1$,
 $$ i(t) \leq 1+ \frac{Ap}{(\alpha -1)\gamma}.$$
 \end{theorem}

\begin{corollary} \label{cor_amp}
In order to contain the malware from propagation, the convergence rate of a learning-based signature generation scheme must have an amplification factor $\alpha > 1$.
\end{corollary}

\begin{corollary} \label{cor_dec}
For $\alpha>1$, we have:
\begin{itemize}
	\item the number of nodes that will be infected is proportional to the deceleration factor $A$.  
	\item substituting $p=\eta k/n$, and $\gamma=\lambda \eta$, gives $i(t) \leq 1+ \frac{A}{(\alpha -1)\lambda}(\frac{k}{n})$ meaning that $i(t)$ is independent from the malware's scanning rate $\eta$.
\end{itemize}

\end{corollary}

 
 \subsection{Model Extensions and Limitations}\label{subsec.gameExtentions}
 
 In this section, we briefly discuss some possible scenarios which require further investigations or can be modeled through the same methodology presented in this work.  Where possible, we provide general guidance for the curious reader on how such cases can be analyzed with a few adjustments to the model.
  
 \textbf{An adaptive adversary with multiple exploits:}
 After observing no progress in the malware propagation mission (i.e., being contained with some limited number of agents), an adaptive adversary may decide to improve and update its attack technology. For instance, a bot herder can update the malware binary on its bots for different purposes. This includes enhancing and extending attack vectors and technologies by adding new functionalities or evasion techniques, migrating to different C\&C servers, and considering recent and revised exploits. Our framework, can also capture such strong and possibly well-funded adversary.
 From the attacker's perspective, each new exploit $E_j$ (or in other words an enhanced attack technology) should be associated with some probability $p_j$ (as each exploit might correspond to a different set of vulnerable nodes within the network). From the defender's perspective, the signature generator, if it cannot generate a single universal signature which matches all the malware's instances, must be able to generate multiple signatures each of which matches some subset of flows in the suspicious flow pool. Therefore, multiple detection rate functions and learning rates $f_j$ should be considered for each class of malware.
 
 \textbf{A deceptive attacker capable of misleading the learning engine:}
There exist various attacks against signature generation schemes especially those with learning based on pattern extraction (see \cite{fogla2006polymorphic,van2007catch,newsome2006paragraph,perdisci2006misleading}). Noise injection attacks in which the attacker can systematically inject noise (or deliberately crafted attack samples) in the training pool to mislead the defender’s learning engine are amongst the most notable challenges for signature generation schemes in adversarial settings. 
In developing the learning-based model, we decided not to consider a \emph{deceptive} adversary since in practice, a malware propagator usually follows a passive attack strategy for the propagation of the malware with a broad audience (e.g., worms/bots), meaning that the attack technology usually remains the same after unleashing the first few samples of the malware. That is the malware propagation game is not a dynamic interaction between the attacker and the defender (especially from the attacker's perspective) and although such attacks are theoretically possible, they have not yet been observed in the wild. 
In general, when dealing with a delusive adversary, the learning rate of the defense system may not always be positive, and the learning engine’s false positive rates should be taken into consideration in the model. We leave this problem for future studies and refer the reader to \cite{venkataraman2008limits} which studies the signature generation in adversarial settings and proves lower bounds on the number of mistakes any pattern extraction learning algorithm can take under common assumptions. 
 
 \textbf{Host-level detection and prevention:} When the malware resides on a machine, it surely exhibits abnormal and erratic behavior (both externals such as unusual port usage \cite{inoue2009automated} and internals such as irregular system call sequences \cite{kolosnjaji2016deep}) on that machine. Therefore, one possibility is taking the knowledge gained at the host-level into consideration for the purpose of attack/malware signature generation by means of monitoring events on endpoint devices through a HIDS. Such signatures can be generated based on the malware residual activity on the target machine or the specific exploit/vulnerability used by the adversary during the infection. Hence, a cumulative function $f_h$ can be considered for the host-level signatures which depends on the total number of so far infected machines. Therefore, one can easily fine-tune the model to capture such defensive scenarios by multiplying the adversary's probability of success at each time step $p_i$ with $1-f_h(i)$ meaning that not only it must hit the right target (i.e., a vulnerable host), the malware should not be neutralized too at the host level. The $f_h$ function gets updated once a new node becomes infected. 
 There exist proposals for combined detection methods, i.e., taking both host-level and network-level information into consideration to fight against a worm/bot malware. For example, \cite{zeng2010detection} introduced a C\&C protocol-independent detection framework based on the combination of information gained from both host and network level bot activities. We decided not to count such strategies, since considering a universal host-level defense system installed on all the network devices is an unrealistic assumption in almost any practical situation.

%% file: sections/num_analysis.tex
The learning-based model is a general malware propagation model with several parameters.  Notice that when the learning function is set to zero, i.e., $f(.)=0$, and $p = \eta \frac{k-i(t)}{n}$, we have exactly the classical epidemic model. Although the effects of cleaning the infection and patching the susceptible population can be easily included in the model, we decided to ignore such conducts since they have been well studied in the literature. Also, we are interested in evaluating the impacts of learning even in the worst case scenarios for the defender (i.e., giving a leg up to the attacker by considering $p= \sup_i p_i$).

For the purpose of illustration, we consider a hypothetical polymorphic worm, propagating in an address space of size $n = 2^{32}$ (the entire IPv4 address space), while the size of susceptible hosts' population is $k=350,000$ . We assume a blind scan strategy for which each infected host performs a uniform scanning in the address space with an average rate of $10,188$ scans per hour\footnote{These are the actual infamous codeRed1v2's epidemic parameters \cite{rohloff2005stochastic}. We use these values only for the purpose of illustrations since the malware is well studied and the parameters are already known.}. Moreover, we consider $i(0) =1$, and $j(0)=0$, meaning that initially there exists one infected machine at time zero, and the defender has no knowledge regarding the attack (signature). 


Although we proved upper and lower bounds on the total number of infected nodes $i(t)$ (see Theorem \ref{theo}), for the general learning-based model, we cannot get closed-form solutions. Instead, we present numerical solutions of the differential equation by using Python \textit{scipy.integrate} package.

\begin{figure}
	\begin{center}
		\scalebox{0.55}
		{\includegraphics{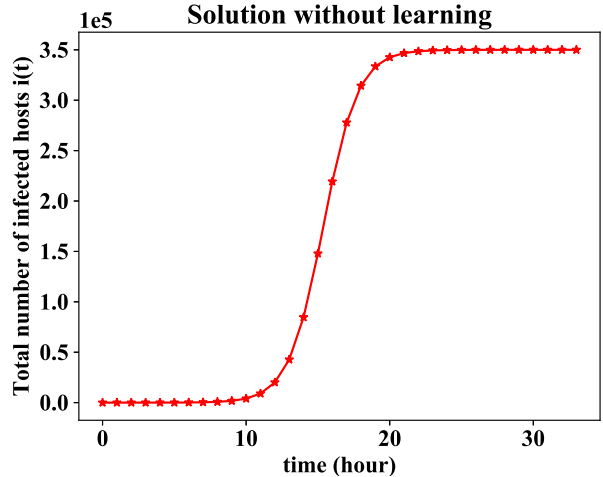}}
		\caption{No learning, classical epidemic model}
		\label{fig.classic}
	\end{center}
\end{figure}

Fig. \ref{fig.classic} depicts the solution of the learning-based model when the learning parameter is set to zero. As mentioned earlier, without a learning process the model is equivalent to the classical epidemic model, and the saturation in the curve depicts how the whole susceptible population will be infected as time elapses.

We now study the impact of different parameters on the total number of infected nodes in the learning-based model.

\begin{figure}
	\begin{center}
		\scalebox{0.55}
		{\includegraphics{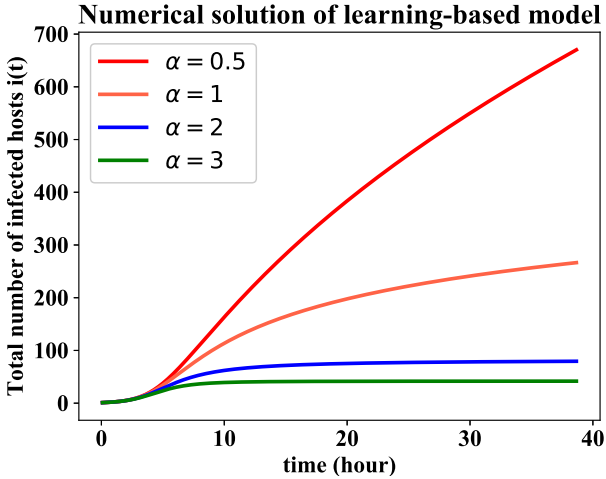}}
		\caption{Numerical solution of learning-based model for $1-f(l) = (1000/(l+1000))^{\alpha}$ while $\lambda=0.001$}
		\label{fig.dif_alpha}
	\end{center}
\end{figure}

Fig. \ref{fig.dif_alpha} depicts how the amplification factor $\alpha$ plays a role in the modeling. The $\Omega(t^{0.5})$ and $\Omega(\ln t)$ growth in the population of infected hosts, for $\alpha=0.5$, and $\alpha=1$ respectively, can be seen from this plot which is consistent with Theorem \ref{theo}. In addition, notice that based on the theorem, for $\alpha>1$, the upper bound on the total number of infected nodes is $i(t)\leq 1+82/(\alpha -1)$, which gives $83$, and $42$ for $\alpha=2$ and $3$ respectively. The value of $i(t)$ for large enough $t$s based on the numerical solution of the differential equation for these two cases is $79.99$ and $41.77$ which is very close to the upper bound. 

\begin{figure}
	\begin{center}
		\scalebox{0.55}
		{\includegraphics{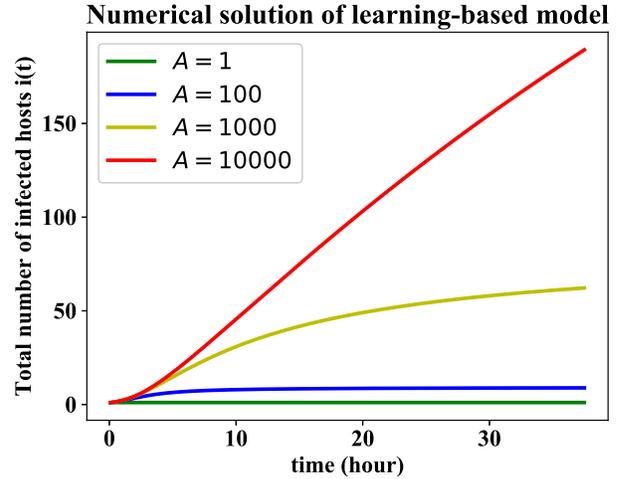}}
		\caption{Numerical solution of learning-based model for $1-f(l) = (A/(l+A))^{2}$ while $\lambda=0.001$}
		\label{fig.dif_a}
	\end{center}
\end{figure}

\begin{figure}
	\begin{center}
		\scalebox{0.55}
		{\includegraphics{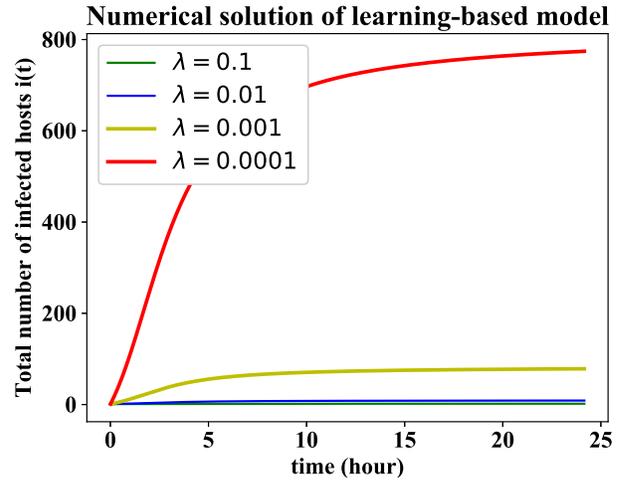}}
		\caption{Numerical solution of learning-based model for $1-f(l) = (1000/(l+1000))^{2}$ and different values of $\lambda$}
		\label{fig.dif_lambda}
	\end{center}
\end{figure}

In order to study the impact of other parameters (i.e., $A$, and $\lambda$) in the learning-based model, we numerically solved the differential equation for a constant $\alpha =2$ to see how the other parameters play a role while the amplification factor is set to an acceptable value. Fig. \ref{fig.dif_a} depicts how the deceleration factor $A$ impacts $i(t)$. In addition, the classifier's accuracy's impact on the malware containment is shown in Fig. \ref{fig.dif_lambda}. As it can be seen from this plot, the larger the $\lambda$, the sooner the attacker's progress is contained in the environment.

%% file: sections/sim.tex
In this section, we want to see how reasonable is the assumption of giving the adversary the advantage of $p= \sup_i p_i$ in our analysis by simulating the malware propagation process. 
Algorithm \ref{alg:sim} provides a high-level overview of malware's self-propagation activity. The simulator's inputs are $\mathcal{N,K}$ (set of all nodes and susceptible nodes within the address space respectively), and an adversarial exploit $\chi$, in addition to the filtering probability $f(l)$ based on the so far collected samples and the corresponding generated signatures, and the defense system's sampling rate $\lambda$. The termination rule is whether the attacker compromises all the susceptible population or the defender finds a true signature, that is $f(l) = 1$. Note that the transmit method returns a tuple, i.e., if the transmitted packet by an infected machine is filtered by the defense system or is sampled at the flow classifier level. 
We simulate two different scenarios. In the first scenario, we give the attacker a leg-up by considering a constant $p$ (i.e., $p = \sup_i p_i = \eta k/n$; this requires removing line \ref{Algline} in Algorithm \ref{alg:sim}, and updating the termination rule in line \ref{termRule} to $|\mathcal{I}| == |\mathcal{K}|$). For the second scenario, we consider $p$ to be a function of so far infected nodes (i.e., $p = \eta(k-i(t))/n$). For both cases, we consider $\lambda = 0.001$ and a learning function of the form $1-f(l) = (\frac{1000}{l+1000})^{2}$ as it satisfies the containment requirement (i.e., $\alpha >1$). We again use the same propagation parameters we used in the numerical analysis section only for the purpose of illustration.

\begin{algorithm}
	\caption{Worm Propagation Simulator}
	\label{alg:sim}
	\begin{algorithmic}[1]
		\Function{Simulator}{$\mathcal{N,K},\chi,\lambda,f(.)$}
		\State {$\mathcal{I} = \emptyset$; $l = 0$;}
		\State {trueSignature = False};
		\State {thoroughInfection = False};
		\While {not trueSignature and not thoroughInfection}
		\State {targetHost $\xleftarrow[\text{}]{\text{\$}} probe(\mathcal{N})$};
		\State {filtered,sampled $\xleftarrow[\text{}]{f(l),\lambda}$$transmit$($\chi$,targetHost)};
		\If {sampled} \label{alg.def_start}
		\State{$l\pluseq 1$};
		\State {update $f(l)$};
		\If {$f(l) == 1$}
		\State {trueSignature = True};
		\State {Output \textit{``Attacker cannot make any progress!''}};
		\EndIf 
		\EndIf \label{alg.def_end}
		\If {targetHost $\in \mathcal{K}$ and not filtered}
		\State $\mathcal{I} = \mathcal{I}~ \cup$ targetHost;
		\State $\mathcal{K} = \mathcal{K}~ \setminus$ targetHost; \label{Algline}
		\If {$|\mathcal{K}| == 0$} \label{termRule}
		\State {thoroughInfection = True};
		\State {Output \textit{``All vulnerable targets are infected!''}};
		\EndIf 
		\EndIf
		\EndWhile
		\EndFunction
	\end{algorithmic}
\end{algorithm}

\begin{figure*}
	\begin{center}
		\scalebox{0.35}
		{\includegraphics{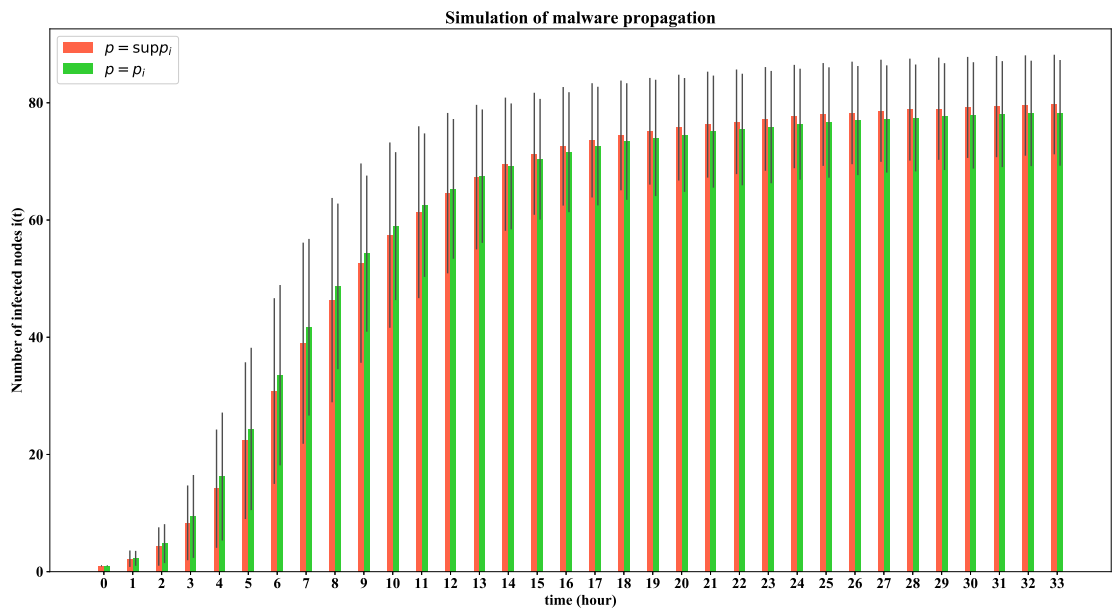}}
		\caption{Simulation of malware propagation for two cases: $p = \sup_i p_i$, and $p = p(i)$}
		\label{fig.Simulations}
	\end{center}
\end{figure*}

\begin{figure}
	\begin{center}
		\scalebox{0.55}
		{\includegraphics{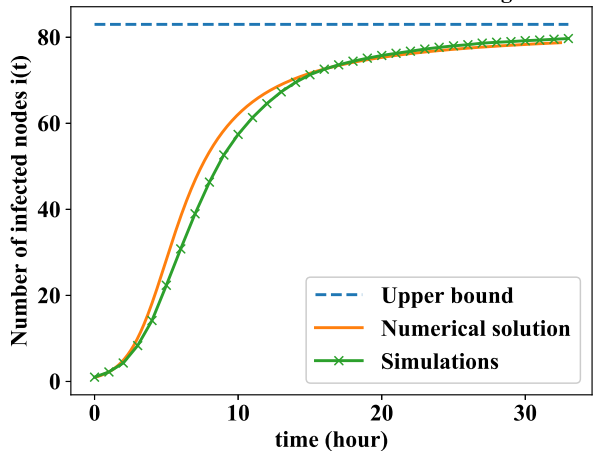}}
		\caption{Comparing the total number of infected nodes based on simulations (using Algorithm \ref{alg:sim} with $p = \sup_i p_i$) and numerical solution of learning-based model for $1-f(l) = (1000/(l+1000))^{2}$ while $\lambda=0.001$}
		\label{fig.comparison}
	\end{center}
\end{figure}

Fig. \ref{fig.Simulations} shows the expected number of infected nodes among 100 simulation runs of the learning-based model. As it can be seen from this figure, the infected population are almost the same for both cases of  $p = \sup_i p_i$, and $p = p(i)$ meaning that $\eta \frac{k-i(t)}{n}\approx \eta k/n =p$ is indeed a good approximation, and considering a constant $p$ in the model is a reasonable assumption. Notice that this behavior was expected since in the learning-based model as time elapses and $f(l)\rightarrow 1$, the chances of making progress for the malware propagator get smaller and smaller (causes the saturation observed in the plots). This means that the value of $i(t)$ remains relatively small in comparison to $k$ and therefore $i(t)$ can be bounded to a number $\ll k$. Fig. \ref{fig.comparison} on the other hand, compares the result of simulations with the numerical solution of the learning-based model. For both scenarios, the upper bound on the size of the infected population can be used as a tight estimate. 
 
%% file: sections/conclusion.tex
This presented work offers a fresh look at a 15-year old research area (worm epidemic modeling). Although the self-propagating class of malware may seem to be an \emph{old}\footnote{The Morris worm was the first computer worm released on the Internet on Nov. 1988, almost three decades ago.} type of a threat due to the decline in the ``worm-like'' common vulnerabilities exposures (CVEs) \cite{cisco:Misc}, the rise of self-propagating ransomware (e.g., the WannaCry cryptoworm) and the highly sophisticated malware such as Stuxnet worm attest that the network vector is one of the few attack vectors which is capable of passing the test of time. 
Taking advantage of the network vector and releasing the malware in a worm-like fashion, meaning that the malware is equipped with a self-propagating functionality, can provide the attackers a more potent malware which is capable of causing widespread damage.
This means that the security community and professionals need to take the self-propagating class of malware more serious to first better understand such malware's life cycle and then to be able to come up with feasible solutions to tackle this problem.
To this end, we revisited the spread of malware phenomenon, especially worm and bot type malware, from an entirely new perspective.  We have modeled the interactions of an adversary and its agents with a defensive system during the construction of a botnet as an incremental online learning process. 
By focusing on the learning rate of a defense system's learning engine and signature generation algorithm, we presented a novel and general propagation model called the \emph{learning-based model} suitable for today's technology and infrastructure. 

Unlike the existing outdated \emph{static} modelings in this area, the learning-based model is a \emph{dynamic} one which can capture the increased knowledge of the defender regarding the attack technology into consideration for bringing next adversarial actions into a halt.
We studied monotonically increasing learning functions for which we showed how different system parameters play a role in the malware containment process.
In particular, we show that a learning function with amplification factor $\alpha \leq 1$ allows the attacker to succeed in its zombie army construction mission. The deceleration factor in learning must remain small enough for the containment purposes meaning that the defender cannot wait for too long learning about a used exploit, and once the learning starts it must continue with an associated detection probability converging fast enough to $1$. Using the learning-based model, we provided a precise bound on the convergence of a learning-based scheme that ensures that the worm propagation reaches minimal saturation in the number of infected hosts during the worm's life cycle. We have shown that learning-based signature generation schemes can be very effective for malware propagation containment purposes only if their convergence rates satisfies our presented criteria. The presented security analysis recommends 
$1-f(l)\leq O(l^{-\alpha})$ with a proof for $\alpha>1$ in our framework, which were consistent with the numerical analysis and simulation results. The attacker needs to find just one exploit for which the defender is too slow to react or too slow in learning. 
Instead of focusing on higher scan rates, the attacker must invest in more stealthier and robust target discovery schemes and malware delivery techniques which would not raise suspicions, leading to smaller value $\lambda$s and therefore fewer attack samples will be provided to the defense system. 

A possible direction for future work is to estimate the learning-based model's parameters, especially the learning rate $f(l)$, based on the size of suspicious traffic pool (i.e., number of observed malicious payloads) given a ``worst-case adversary''. Our framework lays the foundation for such work and allows to provide a worst-case probabilistic bound on the number of compromised nodes based on the estimated $f(l)$ which in turn, offers guidance to the system defender in how to earmark its resources and use the model for damage assessment and prediction purposes.

We believe our study can provide network security researchers and architects valuable lessons and more robust understandings of both attack and defense technologies concerning malware with a broad audience. Besides, with the advent of software-defined networking (SDN) in which the entire network infrastructure can be controlled from a centralized software controller, and the embrace of AutoML in defense technologies, new generations of traditional cyber defense technologies (e.g., HPs, IDPSes, firewalls, etc.), which are more capable, scalable and secure, are already being introduced.
This could be a new opportunity and a fresh start to fight against the botnet phenomenon or the spreading of the malware in general.
For future work, we would like to evaluate the performance of common automatic signature generation schemes using real data from enterprise networks by porting the learning-based containment mechanism to edge and local routers, especially in a software-defined network environment.

%% file: sections/appendix.tex
\section{Proof of Theorem \ref{theo}}\label{sec:appendix}
In order to make the next derivations readable we introduce
\begin{eqnarray*}
y &=& j(t), \\
v &=& y' = j'(t) = \gamma \cdot i(t). 
\end{eqnarray*}
Differential equation (\ref{eqi3}) in terms of $y$ and $v$ after reordering terms reads
\begin{equation} y''  = A^\alpha p \cdot v \cdot (y+A)^{-\alpha}. \label{eqi4} \end{equation}
We derive
$$ y''=v' =  \frac{d v}{dt} = \frac{d v}{d y} \frac{d y}{dt} = y' \frac{d v}{d y} = v \frac{d v}{d y}.$$
Substituting this expression into (\ref{eqi4}), dividing by $v$, and multiplying with $dy$ yields
$$ d v = A^\alpha p \cdot (y+A)^{-\alpha} \cdot d y.$$
Assuming $\alpha\neq 1$, then taking integrals on the left and right side gives the equation
\begin{equation} v = \frac{A^\alpha p}{1-\alpha} \cdot (y+A)^{1-\alpha} + c_1 \label{eqvy} \end{equation}
for some constant $c_1$.
For $t=0$, we have $y(0)=j(0)=0$ and $v(0)=y'(0)= j'(0)= \gamma \cdot i(0)=\gamma$. Substituting this into the above equation solves
\begin{equation} c_1 = \gamma + \frac{Ap}{\alpha -1}. \label{eqc1} \end{equation}

Now we substitute $v=\frac{dy}{dt}$ into (\ref{eqvy}) and multiply both sides with $dt$:
\begin{equation} dy = \left\{ \frac{A^\alpha p}{1-\alpha} \cdot (y+A)^{1-\alpha} + c_1 \right\} \cdot dt. \label{eqde1} \end{equation}
Equivalently, we have
$$ dt = \left\{ \frac{A^\alpha p}{1-\alpha} \cdot (y+A)^{1-\alpha} + c_1 \right\}^{-1} \cdot dy. $$
Taking integrals on the left and right side gives the expression
$$
t = c_2 + \int_{y= y(0)}^{y(t)} \left\{ \frac{A^\alpha p}{1-\alpha} \cdot (y+A)^{1-\alpha} + c_1 \right\}^{-1} dy $$
for some constant $c_2$.
Substituting $t=0$ gives $c_2=0$ and $y(0)=j(0)=0$. Plugging in (\ref{eqc1}) proves
\begin{eqnarray*}
t &=& \int_{y= 0}^{y(t)} \left\{ \frac{A^\alpha p}{1-\alpha} \cdot (y+A)^{1-\alpha} + \gamma + \frac{Ap}{\alpha -1}  \right\}^{-1} dy \\
&=& \frac{\alpha -1}{Ap} \cdot 
 \int_{y= 0}^{y(t)} \left\{ 1- (y/A+1)^{1-\alpha} + \frac{\gamma(\alpha -1)}{Ap}   \right\}^{-1} dy.
 \end{eqnarray*}

 As an immediate consequence  we can take the derivative (using the chain rule) with respect to $t$ on both sides. This shows that (this can also directly be concluded from (\ref{eqvy}))
 \begin{equation} 1 =  \frac{\alpha -1}{Ap} \cdot  \left\{1- (y(t)/A+1)^{1-\alpha} + \frac{\gamma(\alpha -1)}{Ap}   \right\}^{-1}  \cdot y'(t).
 \label{eqdy} \end{equation}
 Notice that $y(t)=j(t)$ is increasing (since $j'(t)=\gamma \cdot i(t)\geq 0$) with $y(t)\geq y(0)=0$. Therefore, if $\alpha> 1$, then  $0\leq (y(t)/A+1)^{1-\alpha}\leq 1$ and
 $$ \left\{1- (y(t)/A+1)^{1-\alpha} + \frac{\gamma(\alpha -1)}{Ap}   \right\}^{-1} \geq \left\{ 1+  \frac{\gamma(\alpha -1)}{Ap}   \right\}^{-1}. $$
 
 \textbf{Case \RNum{1} ($\alpha >1$):} Substituting this into (\ref{eqdy}) yields for $\alpha >1$,
 $$ 1\geq  \frac{\alpha -1}{Ap} \cdot  \left\{ 1+ \frac{\gamma(\alpha -1)}{Ap}   \right\}^{-1} \cdot y'(t)$$
 proving
 $$ i(t)=y'(t)/\gamma \leq  \frac{Ap}{\alpha -1} \cdot  \left\{1+\frac{\gamma(\alpha -1)}{Ap} \right\}/\gamma =1+ \frac{Ap}{(\alpha -1)\gamma}.$$
 
  \textbf{Case \RNum{2} ($\alpha <1$):} As a second consequence of (\ref{eqdy}), we derive for $\alpha<1$,
 \begin{equation} y'(t) = \frac{Ap}{\alpha-1} \left\{ 1- (y(t)/A +1)^{1-\alpha}\right\} + \gamma \geq \gamma=y'(0). \label{lb} \end{equation}
 Hence,
 $$y(t)\geq \gamma\cdot t.$$
 Substituting this back into (\ref{lb}), and noticing that $i(t)= y'(t)/\gamma$ gives
$$i(t) \geq  \frac{Ap}{\alpha-1} \left\{ 1- (\gamma\cdot t/A +1)^{1-\alpha}\right\}/\gamma + 1 = \Omega(t^{1-\alpha}).$$
 This proves that only a learning rate with amplification $\alpha>1$ will prevent the malware from propagating to all $k$ vulnerable nodes.
 
\textbf{Case \RNum{3} ($\alpha=1$):} For completeness, if $\alpha=1$, then taking integrals that led to (\ref{eqvy}) for $\alpha\neq 1$, now lead to
 \begin{equation} \frac{dy}{dt}= v = Ap \ln (y+A) + c_1 \label{eqa1} \end{equation}
 with
 $$ c_1=\gamma - Ap \ln A.$$
 Notice that $y'(t)= Ap \ln (y/A +1) + \gamma \geq \gamma$ and we use the same argument as we did above for $\alpha<1$. We again have $y(t)\geq \gamma \cdot t$ and substituting this back gives
$$i(t) \geq Ap/\gamma \ln (\gamma\cdot t/A +1) +1 = \Omega(\ln t).$$
These lower bounds can be improved by recursively substituting lower bounds back into (\ref{lb}) and (\ref{eqa1}), respectively.